\pdfoutput=1
\documentclass{article}

\PassOptionsToPackage{numbers, compress, sort}{natbib}
\usepackage[preprint]{neurips_2026}

\usepackage[utf8]{inputenc} % allow utf-8 input
\usepackage[T1]{fontenc}    % use 8-bit T1 fonts
\usepackage{amsmath}        % equation, \lVert, etc.
\usepackage{amsfonts}       % blackboard math symbols
\usepackage{booktabs}       % professional-quality tables
\usepackage{enumitem}       % itemize/enumerate spacing options
\usepackage{graphicx}       % \includegraphics with alt text
\graphicspath{{figures/}}
\usepackage{microtype}      % microtypography
\usepackage{xcolor}         % colors (referenced by the style file's macros)
\usepackage{url}            % simple URL typesetting
\usepackage{hyperref}       % hyperlinks; loaded last

\title{Predictor Construction Can Reverse Multimodal Neural Contrasts}

\author{%
  Lucas Nadolskis \\
  University of California Santa Barbara \\
  \texttt{lgilnadolskis@ucsb.edu}
  \And
  Galen Pogoncheff \\
  University of California Santa Barbara \\
  \And
  Michael Beyeler \\
  University of California Santa Barbara
}

\begin{document}

\maketitle

% ===== Abstract =====
% Included from neurips_2025.tex -- do not compile on its own.

\begin{abstract}

Foundation-model features are increasingly used to ask what information neural activity represents, often by comparing prediction gains between nested encoding models.
We show that such multimodal contrasts can change sign when only the conditioning predictor is reconstructed.
Using fMRI from the Natural Scenes Dataset, DINOv2 visual features, and MPNet embeddings of MS COCO captions and Localized Narratives, a caption–narrative contrast in the additional predictive contribution of vision favors narratives when one short caption is compared with a long narrative ($+0.012/{+0.015}$ in Places), but favors captions after approximate word-count matching ($-0.031/{-0.023}$). 
The shift occurs across every measured ROI in both subjects and is driven primarily by differences in language-only prediction.
Comparable contrasts also survive removal of image-specific content-word identity in several ROIs.
These results show that nested neural contrasts do not identify represented content by themselves: predictor construction is part of the experimental design, and matched controls are required for representational claims.

\end{abstract}

\section{Introduction}
\label{intro}

Foundation-model (FM) embeddings are increasingly used as feature spaces for
interpreting neural responses. Vision-model features are compared with neural
responses in model-to-brain benchmarks and encoding analyses
\citep{schrimpf2018brain,conwell2024visual}, while language-model features have
been used to predict or decode cortical activity
\citep{pereira2018toward,huth2016natural}. A common strategy is to add feature
space $A$ to an encoding model that already contains feature space $B$ and
interpret the resulting gain in prediction as information carried by $A$ beyond
$B$. Related contrasts are used to evaluate brain foundation models against
simpler representations \citep{ortegacaro2024brainlm,azabou2023unified} and to
argue that language-derived features explain variance in high-level visual
cortex beyond visual features alone
\citep{wang2023better,doerig2025high,rajaei2026language}.

Interpreting such a gain requires knowing why one predictor outperforms
another. The residual attributed to $A$ depends not only on what $A$
represents, but also on how well the conditioning predictor $B$ explains the
same responses. This matters because alternative predictors can differ in
properties unrelated to the representational distinction of interest, including
input length, lexical composition, annotator count, redundancy, and embedding
construction. Control tasks are therefore central to interpreting learned
probes \citep{hewitt2019designing,feghhi2024what}, while targeted perturbations
can reveal which stimulus properties actually support model-to-brain
correspondence \citep{kauf2024lexical,conwell2024large}. Comparable controls
are less commonly applied to the interpretation of multimodal neural contrasts
themselves.

Here we ask how much visual features add once a text description of the same
image has been accounted for, and whether that residual depends on how the text
predictor is constructed. We use fMRI responses from two extensively sampled
subjects of the Natural Scenes Dataset \citep{allen2022massive}, DINOv2 visual
features, and MPNet embeddings of MS~COCO captions
\citep{lin2014microsoft} and Localized Narratives
\citep{pont2020connecting}. Across comparisons, we hold the image, neural
response, and visual predictor fixed while varying only the language predictor.
This isolates how predictor construction changes the inferred additional
contribution of vision.
Our contributions are threefold:
\begin{itemize}[
  leftmargin=1.25em,
  labelsep=0.45em,
  topsep=0pt,
  itemsep=2pt,
  parsep=0pt,
]
  \item \textbf{We identify a failure mode of nested encoding contrasts:}
  the inferred contribution of vision can change substantially when only the
  conditioning language predictor changes.

  \item \textbf{We show that predictor construction can reverse the result:}
  a contrast between captions and narrative that initially favors narratives reverses after approximate length matching, with the underlying shift appearing across all
  measured ROIs in both subjects.

  \item \textbf{We provide practical identifiability controls:}
  alternative predictor constructions, contrast decomposition, and
  content-word scrambling test whether an observed neural contrast supports
  the representational interpretation assigned to it.
\end{itemize}

\section{Methods}
\label{methods}

All analyses used banded ridge encoding models with a fixed DINOv2 visual
predictor and one MPNet language predictor
\citep{nunez2019voxelwise,latour2022feature}. Only the language predictor
changed across conditions. All feature spaces were projected to 256 dimensions
using \texttt{SparseRandomProjection} \citep{achlioptas2003database}.

\paragraph{Subjects and ROIs}
\label{rois}
We analyzed two extensively sampled subjects from the Natural Scenes Dataset
(NSD) \citep{allen2022massive}. 
Each viewed 9{,}000 unique and 1{,}000 shared images three times; 1.8\,mm
single-trial betas (\texttt{betas\_fithrf\_GLMdenoise\_RR}) were averaged
across repeats. We used subject-native Places (OPA+PPA+RSC)
\citep{epstein1998cortical}, Faces (OFA+FFA-1/2)
\citep{kanwisher1997fusiform}, Bodies (EBA+FBA-1/2) and EBA
\citep{downing2001cortical}, as well as V1 and V4.
EBA is a subset of Bodies and was not treated as independent in pairwise
comparisons.

Within each ROI, we retained voxels with NSD noise-corrected SNR
(ncsnr) $\geq 0.2$ and used the same voxels for every model in a contrast.
Results were stable at thresholds of $0.3$ and $0.4$; ROI definitions, voxel
counts, and threshold sensitivity are reported in
Appendix~\ref{app:rois}.

\paragraph{Features and text conditions}
\label{textdata}

We paired DINOv2 and MPNet by encoder depth: layer 2 for V1, layer 6 for V4,
and the final blocks (DINOv2 11; MPNet 12) for the category-selective ROIs
\citep{oquab2024dinov2,song2020mpnet}. ROI comparisons were restricted to
matching layer pairs; extraction details are in Appendix~\ref{app:embeddings}.

MS~COCO captions averaged 10.4 words
\citep{lin2014microsoft}; we used one caption or concatenated four or five,
with four averaging 41.7 words. Localized Narratives averaged 41.3 words
\citep{pont2020connecting}. Thus, four captions approximately matched a
narrative in length, though not in annotator count or text structure. We also
truncated narratives to each image's caption length, retaining either the
opening words or an interior span. The matched four-caption and narrative
conditions contained 48.1 and 47.0 MPNet tokens on average. All conditions are
listed in Table~\ref{tab:tab_predictors}; further details are in
Appendix~\ref{app:embeddings}.

\paragraph{Scramble control}
\label{subsec:scrambles}
To test whether the contrasts depended on image-specific lexical content, we
constructed scrambled versions of both text datasets using spaCy
\citep{honnibal2020spacy}. Nouns, proper nouns, verbs, adjectives, and numerals
were replaced with different words drawn from the same dataset, matched on
part of speech when possible, and sampled independently of the image. Replacement
words could not occur as content words in the original description.

Following prior lexical-semantic perturbation controls
\citep{kauf2024lexical,hewitt2019designing}, the scramble preserved text
length, function words, part-of-speech structure, and content-word positions
while removing the original image-specific content words. A reconstruction
control verified that tokenization and reconstruction alone did not affect the
results (Appendix~\ref{app:scrambling}).

\paragraph{Encoding model and contrast}
\label{subsec:model}
We fit banded ridge models with separate penalties for the vision and language
bands \citep{nunez2019voxelwise,latour2022feature}. Penalties were selected per
voxel by five-fold nested cross-validation; the regularization grid and
implementation details are given in Appendix~\ref{app:ridge}.

Prediction accuracy, $r$, was the Pearson correlation between out-of-fold
predictions and repeat-averaged responses, normalized by each voxel's noise
ceiling \citep{schoppe2016measuring,allen2022massive} and then averaged within
ROI.

For each language predictor $L$, we fit a joint vision--language model and a
language-only model. We define the additional predictive contribution of vision
as
\begin{equation}
    \Delta_V(L)=r(V{+}L)-r(L),
    \label{eq:uniquev}
\end{equation}
where $r$ is the noise-ceiling-normalized Pearson correlation defined above.

A caption--narrative comparison contrasts this quantity for two language
predictors, $L_A$ and $L_B$, while holding the images, neural responses, and
visual predictor fixed:
\begin{equation}
  \Delta_V(L_A) - \Delta_V(L_B)
  =
  \left[r(V{+}L_A)-r(V{+}L_B)\right]
  -
  \left[r(L_A)-r(L_B)\right].
  \label{eq:contrast}
\end{equation}
This decomposition separates changes in joint vision--language prediction from
changes in language-only prediction. Absolute accuracies for all conditions are
reported in Appendix Table~\ref{tab:results}.

\paragraph{Inference}
\label{subsec:robustness}

Uncertainty was estimated separately for each subject using a paired bootstrap
over the 10{,}000 evaluation images
\citep{efron1979bootstrap,efron1994introduction}. For each of 1{,}000 draws,
the same resampled image indices were applied to every model entering a
contrast; models were not refit. We report full-sample point estimates with
95\% percentile bootstrap intervals.

Because only two subjects were analyzed, we report subject-specific effects
rather than population-level inference.

\section{Results}
\label{results}

\paragraph{The importance of text length}
\label{subsec:length}
In Places, the original one-caption versus Localized Narrative contrast was
positive: $+0.0118$ $[+0.0066,+0.0166]$ in subj01 and
$+0.0147$ $[+0.0092,+0.0196]$ in subj02. Thus, the narrative left less
additional contribution for vision, but it was also roughly four times longer
than a caption.

\begin{figure}[!t]
  \centering
  \includegraphics[width=\linewidth,
    alt={Three-panel figure showing how predictor construction changes multimodal neural contrasts. Panel A plots the additional contribution of vision against text length in Places for two subjects: longer text generally reduces the vision contribution, and the caption–narrative contrast reverses after approximate length matching. Panel B shows that this length-matching shift occurs in the same direction across all six ROIs and both subjects, with every matched-length contrast becoming negative. Panel C decomposes the Places contrast and shows that changes in language-only prediction are larger than changes in the joint vision–language model in both subjects and both comparisons.}
  ]{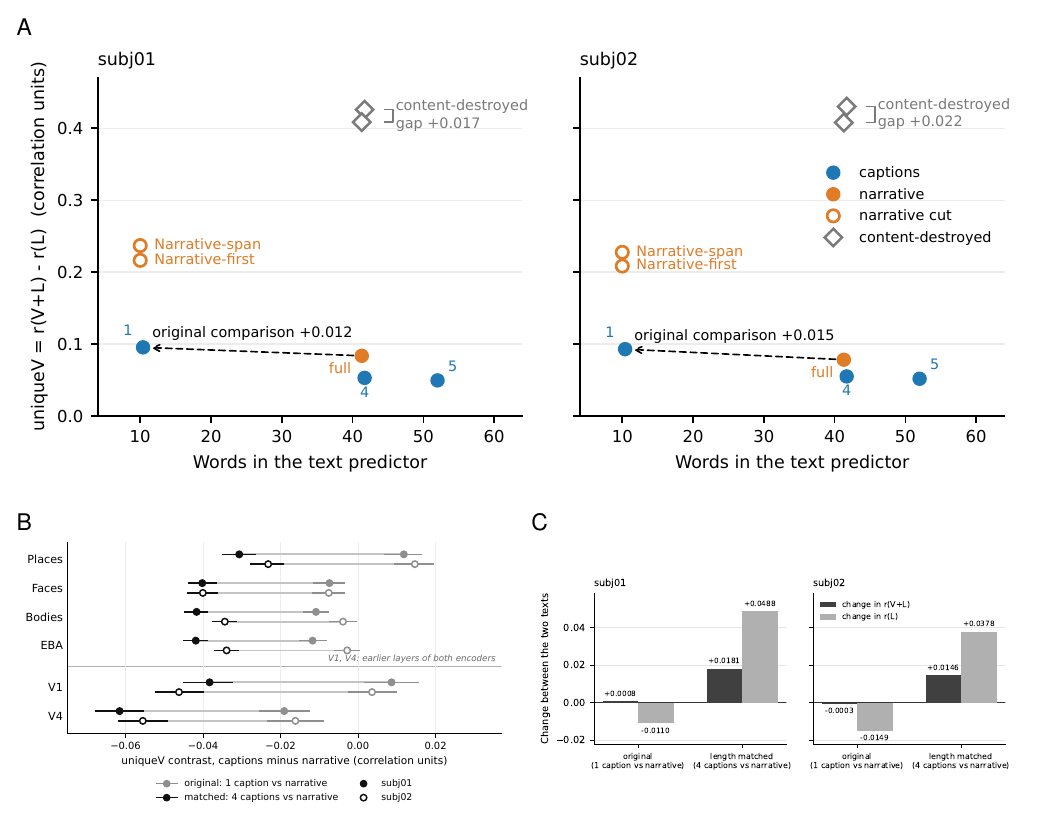}
  \caption{Predictor construction changes multimodal neural contrasts.
    \textbf{(A)} In Places, $\Delta_V(L)$ decreases with text length. The original
    one-caption versus narrative contrast (dashed arrow) favors the narrative but
    reverses when four captions approximately match narrative length. Open orange
    points are caption-length narrative truncations; grey diamonds are
    content-destroyed controls.
    \textbf{(B)} Length matching shifts every ROI in both subjects toward the
    caption predictor. Only Places and V1 in subj01 cross zero, but all
    matched-length contrasts are negative. Filled/open markers indicate
    subj01/subj02; error bars are 95\% bootstrap intervals.
    \textbf{(C)} Decomposition of the Places contrast into changes in joint
    prediction, $r(V{+}L)$, and language-only prediction, $r(L)$. The
    language-only term is larger in magnitude in both subjects and comparisons.}
  \label{fig:main}
\end{figure}

Figure~\ref{fig:main}A shows that $\Delta_V(L)$ decreased with text length for
both predictor families. When four concatenated captions (41.7 words) were
matched to a narrative (41.3 words), the contrast reversed to
$-0.0307$ $[-0.0351,-0.0262]$ and
$-0.0232$ $[-0.0278,-0.0192]$. Truncating narratives to caption length gave
the same result: one caption versus Narrative-first yielded
$-0.121/{-0.116}$, and versus Narrative-span
$-0.141/{-0.135}$, for subj01/subj02. The two narrative truncations themselves
differed by $0.020/0.019$, showing that predictor construction matters even at
fixed word count.

Thus, captions left less additional contribution for vision when length was
matched either by shortening narratives or combining captions. On the
1{,}000 images shared across subjects, the matched-length contrast remained
negative in Faces, Bodies, and EBA but was less consistent in Places
(Appendix~\ref{app:shared_stimuli}).

\paragraph{Length matching shifts every region in the same direction}
\label{subsec:regions}

Figure~\ref{fig:main}B compares the original one-caption contrast with the
matched-length four-caption contrast across six ROIs. Before matching, Places
was the only category-selective ROI with a positive contrast
($+0.012/{+0.015}$); Faces and Bodies were already negative, and EBA was
negative or near zero. V1 was positive only in subj01, while V4 was negative
in both subjects.

Length matching shifted every ROI toward the caption predictor by
$-0.030$ to $-0.050$, with all shifts excluding zero. After matching, every
contrast was negative and every bootstrap interval excluded zero. Sign reversal
therefore occurred only where narratives initially led, in Places for both
subjects and V1 for subj01, but the shift itself was consistent across ROIs.

Among the category-selective ROIs, the matched-length contrast was least
negative in Places; its pairwise differences from Faces and Bodies excluded
zero. EBA was not tested independently because it is a subset of Bodies.
V1 and V4 used different DINOv2--MPNet layer pairs and were not compared
quantitatively with the category-selective ROIs.

\paragraph{Most of each contrast is driven by language-only prediction}
\label{subsec:mechanism}

Equation~\ref{eq:contrast} separates each caption--narrative contrast into
changes in joint vision--language prediction and language-only prediction
(Figure~\ref{fig:main}C).

In Places, the original one-caption contrast was almost entirely language-only:
the joint term was $+0.0008/{-0.0003}$ for subj01/subj02, versus
$-0.0110/{-0.0149}$ for the language-only term. After length matching, both
terms favored captions, but the language-only term remained larger
($+0.0488/{+0.0378}$ versus $+0.0181/{+0.0146}$).

This pattern held across ROIs. The language-only term was larger in magnitude
than the joint term in every ROI, subject, and comparison; for the matched
condition it was 2.2--3.1 times larger in category-selective cortex and
14--24 times larger in V1/V4 (Table~\ref{tab:decomposition}). Thus, the
contrast is driven primarily by the conditioning language predictor, although
the joint model also changes after matching.

\paragraph{Contrasts persist after removing image-specific lexical content}
\label{subsec:floor}

We repeated the matched-length comparison after scrambling content words
(Section~\ref{subsec:scrambles}). In Faces, Bodies, and EBA, the scrambled
contrast remained negative
($-0.062/{-0.052}$, $-0.081/{-0.038}$, and $-0.084/{-0.045}$),
showing that comparable effects can survive removal of the original
image-specific lexical content. Magnitude was somewhat aggregation-dependent
for Bodies and EBA in subj02 (Table~\ref{tab:aggregation}).

The scramble preserves several description-level properties that can themselves
predict neural responses; for example, determiner count alone reached
$r(L)=0.14/0.16$ in Places
(Appendix~\ref{app:scramble_extra}). The surviving contrasts therefore cannot
be uniquely attributed to the lexical-semantic content removed by the
manipulation.

Places differed: the scrambled contrast was
$+0.017/{+0.022}$, opposite to the intact matched-length contrast
($-0.031/{-0.023}$; Figure~\ref{fig:main}A). Because scrambling also
substantially reduced language-model accuracy, we treat it as an
identifiability control rather than as a separable component of the intact
contrast (Table~\ref{tab:results}).

\section{Discussion}
\label{sec:discussion}

\paragraph{Predictor construction.}
We asked how the construction of a conditioning language predictor affects the
estimated additional contribution of vision in multimodal encoding models.
A caption--narrative contrast that initially favored narratives reversed after
approximate word-count matching, and the shift occurred across every measured
ROI in both subjects. Decomposing the contrast showed that most of its magnitude
came from differences in language-only prediction rather than from differences
between the joint vision--language models. The joint term was nevertheless
nonzero after matching. The main point is therefore not that the contrast is
entirely artifactual, but that its representational interpretation depends on
how the conditioning predictor is built.

\paragraph{Length matching.}
The original comparison confounded description type with a roughly fourfold
difference in text length. Matching word count reversed its sign, so the
original effect was not stable to this basic control. But concatenating four
captions also changes annotator count, redundancy, lexical diversity, and other
properties of the predictor. We therefore do not interpret the reversal as a
pure causal effect of length. The narrative truncations make the broader point:
even within the same corpus, with word count matched image by image, changing
which portion of the text is embedded changes $\Delta_V(L)$. Predictor
construction is therefore part of the experimental design rather than a neutral
preprocessing choice.

\paragraph{Mechanism.}
The Appendix analyses further constrain the length effect. Random 10-, 20-, and
41-word subsamples of the same narratives produced monotonically decreasing
$\Delta_V(L)$ in Places as more words were retained
(Appendix~\ref{app:subsample}), and the random 10-word condition fell in the
same regime as the contiguous caption-length cuts. Averaging embeddings from
multiple independent 10-word views recovered roughly three quarters of the gap
between a single 10-word view and the full narrative, even though no encoder
pass contained more than 10 words.

This pattern is compatible with reduced estimator variance, greater content
coverage, or both. Averaging more short views both stabilizes the representation
and samples more of the narrative, so the analysis does not isolate one
mechanism. Within the full-narrative corpus, longer descriptions also left less
additional contribution for vision across all ROIs
(Appendix~\ref{app:dose}), while embedding norm explained none of the effect.
The dependence on text length is therefore not explained by sequence position
or embedding magnitude alone.

\paragraph{Scrambling.}
After image-specific content words were replaced, contrasts comparable to the
intact effect remained in Faces, Bodies, and EBA. In these ROIs, the intact
caption--narrative contrast cannot be uniquely attributed to the
lexical-semantic content removed by the scramble. Places behaved differently:
the scrambled contrast reversed sign relative to the intact matched-length
contrast, so the properties preserved by the manipulation were not sufficient
to explain the caption advantage there. Because scrambling preserves several
description-level properties while substantially changing overall prediction
accuracy, it is best treated as an identifiability control rather than as a
separable estimate of ``form'' or ``content''.

\paragraph{Limitations and implications.}
Our conclusions are limited to two extensively sampled NSD subjects and to the
DINOv2--MPNet feature spaces studied here. The approximately length-matched
caption condition also differs from the narrative in annotator count and other
text properties, so these experiments do not isolate a single cause of the
reversal. Nor do they imply that language-derived features fail to capture
meaningful neural representations.

The practical implication is narrower. Nested contrasts should not be
interpreted from the difference alone. Obvious nuisance dimensions should be
matched where possible, the component prediction terms should be inspected,
and the result should be tested under alternative predictor constructions or
targeted perturbations. Differences between embedding-based predictors are not,
by themselves, sufficient to identify what information a brain region
represents.

% Acknowledgments.  A plain starred section is used instead of the template's
% \begin{ack} environment (which depends on the environ package, unsupported by
% LaTeXML).  Uncomment and fill in before uploading.
% \section*{Acknowledgments and Disclosure of Funding}
% ...

% ===== References =====
% Included from neurips_2025.tex -- do not compile on its own.
\clearpage

\bibliographystyle{plainnat}
\bibliography{refs}

\begin{thebibliography}{38}
\providecommand{\natexlab}[1]{#1}
\providecommand{\url}[1]{\texttt{#1}}
\expandafter\ifx\csname urlstyle\endcsname\relax
  \providecommand{\doi}[1]{doi: #1}\else
  \providecommand{\doi}{doi: \begingroup \urlstyle{rm}\Url}\fi

\bibitem[Achlioptas(2003)]{achlioptas2003database}
Dimitris Achlioptas.
\newblock Database-friendly random projections: Johnson-lindenstrauss with
  binary coins.
\newblock \emph{Journal of Computer and System Sciences}, 66\penalty0
  (4):\penalty0 671--687, 2003.

\bibitem[Allen et~al.(2022)Allen, St-Yves, Wu, Breedlove, Prince, Dowdle, Nau,
  Caron, Pestilli, Charest, Hutchinson, Naselaris, and Kay]{allen2022massive}
Emily~J. Allen, Ghislain St-Yves, Yihan Wu, Jesse~L. Breedlove, Jacob~S.
  Prince, Logan~T. Dowdle, Matthias Nau, Brad Caron, Franco Pestilli, Ian
  Charest, J.~Benjamin Hutchinson, Thomas Naselaris, and Kendrick Kay.
\newblock A massive 7t fmri dataset to bridge cognitive neuroscience and
  artificial intelligence.
\newblock \emph{Nature Neuroscience}, 25:\penalty0 116--126, 2022.

\bibitem[Azabou et~al.(2023)Azabou, Arora, Ganesh, Mao, Nachimuthu, Mendelson,
  et~al.]{azabou2023unified}
Mehdi Azabou, Vinam Arora, Venkataramana Ganesh, Ximeng Mao, Santosh
  Nachimuthu, Michael~J. Mendelson, et~al.
\newblock A unified, scalable framework for neural population decoding.
\newblock In \emph{Advances in Neural Information Processing Systems}, 2023.

\bibitem[Conwell et~al.(2024{\natexlab{a}})Conwell, MacMahon, Vinken, Sharma,
  Jagadeesh, Prince, Alvarez, Konkle, Isik, and Livingstone]{conwell2024visual}
Colin Conwell, Emalie MacMahon, Kasper Vinken, Saloni Sharma, Akshay Jagadeesh,
  Jacob~S. Prince, George~A. Alvarez, Talia Konkle, Leyla Isik, and Margaret
  Livingstone.
\newblock Is visual cortex really ``language-aligned''? perspectives from
  model-to-brain comparisons in human and monkeys on the natural scenes
  dataset.
\newblock In \emph{Conference on Cognitive Computational Neuroscience (CCN)},
  2024{\natexlab{a}}.

\bibitem[Conwell et~al.(2024{\natexlab{b}})Conwell, Prince, Kay, Alvarez, and
  Konkle]{conwell2024large}
Colin Conwell, Jacob~S. Prince, Kendrick~N. Kay, George~A. Alvarez, and Talia
  Konkle.
\newblock A large-scale examination of inductive biases shaping high-level
  visual representation in brains and machines.
\newblock \emph{Nature Communications}, 15, 2024{\natexlab{b}}.

\bibitem[Doerig et~al.(2025)Doerig, Kietzmann, Allen, Wu, Naselaris, Kay, and
  Charest]{doerig2025high}
Adrien Doerig, Tim~C. Kietzmann, Emily Allen, Yihan Wu, Thomas Naselaris,
  Kendrick Kay, and Ian Charest.
\newblock High-level visual representations in the human brain are aligned with
  large language models.
\newblock \emph{Nature Machine Intelligence}, 7, 2025.

\bibitem[Downing et~al.(2001)Downing, Jiang, Shuman, and
  Kanwisher]{downing2001cortical}
Paul~E. Downing, Yuhong Jiang, Miles Shuman, and Nancy Kanwisher.
\newblock A cortical area selective for visual processing of the human body.
\newblock \emph{Science}, 293:\penalty0 2470--2473, 2001.

\bibitem[Dupr{\'e}~la Tour et~al.(2022)Dupr{\'e}~la Tour, Eickenberg,
  Nunez-Elizalde, and Gallant]{latour2022feature}
Tom Dupr{\'e}~la Tour, Michael Eickenberg, Anwar~O. Nunez-Elizalde, and Jack~L.
  Gallant.
\newblock Feature-space selection with banded ridge regression.
\newblock \emph{NeuroImage}, 264:\penalty0 119728, 2022.

\bibitem[Efron(1979)]{efron1979bootstrap}
Bradley Efron.
\newblock Bootstrap methods: Another look at the jackknife.
\newblock \emph{The Annals of Statistics}, 7\penalty0 (1):\penalty0 1--26,
  1979.
\newblock \doi{10.1214/aos/1176344552}.

\bibitem[Efron and Tibshirani(1994)]{efron1994introduction}
Bradley Efron and Robert~J. Tibshirani.
\newblock \emph{An Introduction to the Bootstrap}.
\newblock Chapman \& Hall/CRC, Boca Raton, FL, 1994.

\bibitem[Epstein and Kanwisher(1998)]{epstein1998cortical}
Russell Epstein and Nancy Kanwisher.
\newblock A cortical representation of the local visual environment.
\newblock \emph{Nature}, 392:\penalty0 598--601, 1998.

\bibitem[Fedorenko et~al.(2010)Fedorenko, Hsieh, Nieto-Casta{\~n}{\'o}n,
  Whitfield-Gabrieli, and Kanwisher]{fedorenko2010new}
Evelina Fedorenko, Po-Jang Hsieh, Alfonso Nieto-Casta{\~n}{\'o}n, Susan
  Whitfield-Gabrieli, and Nancy Kanwisher.
\newblock New method for {fMRI} investigations of language: defining {ROIs}
  functionally in individual subjects.
\newblock \emph{Journal of Neurophysiology}, 104\penalty0 (2):\penalty0
  1177--1194, 2010.

\bibitem[Feghhi et~al.(2024)Feghhi, Hadidi, Song, Blank, and
  Kao]{feghhi2024what}
Ebrahim Feghhi, Nima Hadidi, Bryan Song, Idan~A. Blank, and Jonathan~C. Kao.
\newblock What are large language models mapping to in the brain? a case
  against over-reliance on brain scores.
\newblock \emph{arXiv preprint arXiv:2406.01538}, 2024.

\bibitem[Hewitt and Liang(2019)]{hewitt2019designing}
John Hewitt and Percy Liang.
\newblock Designing and interpreting probes with control tasks.
\newblock In \emph{Proceedings of EMNLP-IJCNLP}, pages 2733--2743, 2019.

\bibitem[Honnibal et~al.(2020)Honnibal, Montani, Van~Landeghem, and
  Boyd]{honnibal2020spacy}
Matthew Honnibal, Ines Montani, Sofie Van~Landeghem, and Adriane Boyd.
\newblock {spaCy}: Industrial-strength natural language processing in {Python},
  2020.

\bibitem[Huth et~al.(2016)Huth, de~Heer, Griffiths, Theunissen, and
  Gallant]{huth2016natural}
Alexander~G. Huth, Wendy~A. de~Heer, Thomas~L. Griffiths, Fr{\'e}d{\'e}ric~E.
  Theunissen, and Jack~L. Gallant.
\newblock Natural speech reveals the semantic maps that tile human cerebral
  cortex.
\newblock \emph{Nature}, 532:\penalty0 453--458, 2016.

\bibitem[Kanwisher et~al.(1997)Kanwisher, McDermott, and
  Chun]{kanwisher1997fusiform}
Nancy Kanwisher, Josh McDermott, and Marvin~M. Chun.
\newblock The fusiform face area: a module in human extrastriate cortex
  specialized for face perception.
\newblock \emph{Journal of Neuroscience}, 17\penalty0 (11):\penalty0
  4302--4311, 1997.

\bibitem[Kauf et~al.(2024)Kauf, Tuckute, Levy, Andreas, and
  Fedorenko]{kauf2024lexical}
Carina Kauf, Greta Tuckute, Roger Levy, Jacob Andreas, and Evelina Fedorenko.
\newblock Lexical-semantic content, not syntactic structure, is the main
  contributor to {ANN}-brain similarity of {fMRI} responses in the language
  network.
\newblock \emph{Neurobiology of Language}, 5\penalty0 (1):\penalty0 7--42,
  2024.

\bibitem[Lerner et~al.(2011)Lerner, Honey, Silbert, and
  Hasson]{lerner2011topographic}
Yulia Lerner, Christopher~J. Honey, Lauren~J. Silbert, and Uri Hasson.
\newblock Topographic mapping of a hierarchy of temporal receptive windows
  using a narrated story.
\newblock \emph{Journal of Neuroscience}, 31\penalty0 (8):\penalty0 2906--2915,
  2011.

\bibitem[Lin et~al.(2014)Lin, Maire, Belongie, Hays, Perona, Ramanan,
  Doll{\'a}r, and Zitnick]{lin2014microsoft}
Tsung-Yi Lin, Michael Maire, Serge Belongie, James Hays, Pietro Perona, Deva
  Ramanan, Piotr Doll{\'a}r, and C.~Lawrence Zitnick.
\newblock Microsoft {COCO}: Common objects in context.
\newblock In \emph{European Conference on Computer Vision (ECCV)}, pages
  740--755, 2014.
\newblock \doi{10.1007/978-3-319-10602-1\_48}.

\bibitem[Miller and Isard(1963)]{miller1963some}
George~A. Miller and Stephen Isard.
\newblock Some perceptual consequences of linguistic rules.
\newblock \emph{Journal of Verbal Learning and Verbal Behavior}, 2\penalty0
  (3):\penalty0 217--228, 1963.

\bibitem[Mollica et~al.(2020)Mollica, Siegelman, Diachek, Piantadosi, Mineroff,
  Futrell, Kean, Qian, and Fedorenko]{mollica2020composition}
Francis Mollica, Matthew Siegelman, Evgeniia Diachek, Steven~T. Piantadosi,
  Zachary Mineroff, Richard Futrell, Hope Kean, Peng Qian, and Evelina
  Fedorenko.
\newblock Composition is the core driver of the language-selective network.
\newblock \emph{Neurobiology of Language}, 1\penalty0 (1):\penalty0 104--134,
  2020.

\bibitem[Naselaris et~al.(2021)Naselaris, Allen, and
  Kay]{naselaris2021extensive}
Thomas Naselaris, Emily Allen, and Kendrick Kay.
\newblock Extensive sampling of single cortical subjects.
\newblock \emph{Current Opinion in Behavioral Sciences}, 40:\penalty0 45--51,
  2021.
\newblock \doi{10.1016/j.cobeha.2020.12.008}.

\bibitem[Nunez-Elizalde et~al.(2019)Nunez-Elizalde, Huth, and
  Gallant]{nunez2019voxelwise}
Anwar~O. Nunez-Elizalde, Alexander~G. Huth, and Jack~L. Gallant.
\newblock Voxelwise encoding models with non-spherical multivariate normal
  priors.
\newblock \emph{NeuroImage}, 197:\penalty0 482--492, 2019.

\bibitem[O'Connor and Andreas(2021)]{oconnor2021context}
Joe O'Connor and Jacob Andreas.
\newblock What context features can transformer language models use?
\newblock In \emph{Proceedings of ACL-IJCNLP}, pages 851--864, 2021.

\bibitem[Oquab et~al.(2024)Oquab, Darcet, Moutakanni, Vo, Szafraniec, Khalidov,
  et~al.]{oquab2024dinov2}
Maxime Oquab, Timoth{\'e}e Darcet, Th{\'e}o Moutakanni, Huy Vo, Marc
  Szafraniec, Vasil Khalidov, et~al.
\newblock {DINOv2}: Learning robust visual features without supervision.
\newblock \emph{Transactions on Machine Learning Research}, 2024.

\bibitem[Ortega~Caro et~al.(2024)Ortega~Caro, Oliveira~Fonseca, Averill, Rizvi,
  Rosati, Cross, et~al.]{ortegacaro2024brainlm}
Josue Ortega~Caro, Antonio~H. Oliveira~Fonseca, Christopher Averill, Syed~A.
  Rizvi, Matteo Rosati, James~L. Cross, et~al.
\newblock {BrainLM}: A foundation model for brain activity recordings.
\newblock In \emph{International Conference on Learning Representations}, 2024.

\bibitem[Pereira et~al.(2018)Pereira, Lou, Pritchett, Ritter, Gershman,
  Kanwisher, Botvinick, and Fedorenko]{pereira2018toward}
Francisco Pereira, Bin Lou, Brianna Pritchett, Samuel Ritter, Samuel~J.
  Gershman, Nancy Kanwisher, Matthew Botvinick, and Evelina Fedorenko.
\newblock Toward a universal decoder of linguistic meaning from brain
  activation.
\newblock \emph{Nature Communications}, 9:\penalty0 963, 2018.

\bibitem[Pont-Tuset et~al.(2020)Pont-Tuset, Uijlings, Changpinyo, Soricut, and
  Ferrari]{pont2020connecting}
Jordi Pont-Tuset, Jasper Uijlings, Soravit Changpinyo, Radu Soricut, and
  Vittorio Ferrari.
\newblock Connecting vision and language with localized narratives.
\newblock In \emph{European Conference on Computer Vision (ECCV)}, 2020.

\bibitem[Rajaei et~al.(2026)Rajaei, Afshar, Cichy, and
  Soltanian-Zadeh]{rajaei2026language}
Karim Rajaei, Arian Afshar, Radoslaw~Martin Cichy, and Hamid Soltanian-Zadeh.
\newblock Language-aligned models and structured scene descriptions reveal
  sensitivity to compositional scene structure in the high-level visual cortex.
\newblock \emph{bioRxiv}, pages 2026--07, 2026.

\bibitem[Reimers and Gurevych(2019)]{reimers2019sentence}
Nils Reimers and Iryna Gurevych.
\newblock Sentence-{BERT}: Sentence embeddings using siamese {BERT}-networks.
\newblock In \emph{Proceedings of EMNLP-IJCNLP}, 2019.

\bibitem[Schoppe et~al.(2016)Schoppe, Harper, Willmore, King, and
  Schnupp]{schoppe2016measuring}
Oliver Schoppe, Nicol~S. Harper, Ben D.~B. Willmore, Andrew~J. King, and Jan
  W.~H. Schnupp.
\newblock Measuring the performance of neural models.
\newblock \emph{Frontiers in Computational Neuroscience}, 10:\penalty0 10,
  2016.
\newblock \doi{10.3389/fncom.2016.00010}.

\bibitem[Schrimpf et~al.(2018)Schrimpf, Kubilius, Hong, Majaj, Rajalingham,
  Issa, et~al.]{schrimpf2018brain}
Martin Schrimpf, Jonas Kubilius, Ha~Hong, Najib~J. Majaj, Rishi Rajalingham,
  Elias~B. Issa, et~al.
\newblock Brain-score: Which artificial neural network for object recognition
  is most brain-like?
\newblock \emph{bioRxiv 407007}, 2018.

\bibitem[Sinha et~al.(2021)Sinha, Jia, Hupkes, Pineau, Williams, and
  Kiela]{sinha2021masked}
Koustuv Sinha, Robin Jia, Dieuwke Hupkes, Joelle Pineau, Adina Williams, and
  Douwe Kiela.
\newblock Masked language modeling and the distributional hypothesis: Order
  word matters pre-training for little.
\newblock In \emph{Proceedings of EMNLP}, pages 2888--2913, 2021.

\bibitem[Song et~al.(2020)Song, Tan, Qin, Lu, and Liu]{song2020mpnet}
Kaitao Song, Xu~Tan, Tao Qin, Jianfeng Lu, and Tie-Yan Liu.
\newblock {MPNet}: Masked and permuted pre-training for language understanding.
\newblock In \emph{Advances in Neural Information Processing Systems}, 2020.

\bibitem[Steegen et~al.(2016)Steegen, Tuerlinckx, Gelman, and
  Vanpaemel]{steegen2016increasing}
Sara Steegen, Francis Tuerlinckx, Andrew Gelman, and Wolf Vanpaemel.
\newblock Increasing transparency through a multiverse analysis.
\newblock \emph{Perspectives on Psychological Science}, 11\penalty0
  (5):\penalty0 702--712, 2016.
\newblock \doi{10.1177/1745691616658637}.

\bibitem[Wang et~al.(2023)Wang, Kay, Naselaris, Tarr, and
  Wehbe]{wang2023better}
Aria~Y. Wang, Kendrick Kay, Thomas Naselaris, Michael~J. Tarr, and Leila Wehbe.
\newblock Better models of human high-level visual cortex emerge from natural
  language supervision with a large and diverse dataset.
\newblock \emph{Nature Machine Intelligence}, 5:\penalty0 1415--1426, 2023.

\bibitem[Yuksekgonul et~al.(2023)Yuksekgonul, Bianchi, Kalluri, Jurafsky, and
  Zou]{yuksekgonul2023when}
Mert Yuksekgonul, Federico Bianchi, Pratyusha Kalluri, Dan Jurafsky, and James
  Zou.
\newblock When and why vision-language models behave like bags-of-words, and
  what to do about it?
\newblock In \emph{International Conference on Learning Representations}, 2023.

\end{thebibliography}

% ===== Appendix =====
% Included from neurips_2025.tex -- do not compile on its own.

\clearpage
\appendix

\setcounter{table}{0}
\renewcommand{\thetable}{A\arabic{table}}

\setcounter{figure}{0}
\renewcommand{\thefigure}{A\arabic{figure}}

% hyperref anchor names (\theH...) must stay unique between main-text and
% appendix floats.  Guarded because only hyperref defines them; arXiv's HTML
% converter (LaTeXML) does not.
\makeatletter
\@ifundefined{theHtable}{}{\renewcommand{\theHtable}{A\arabic{table}}}
\@ifundefined{theHfigure}{}{\renewcommand{\theHfigure}{A\arabic{figure}}}
\makeatother

\section{Additional Methods and Analyses}
\label{app:methods}

% ---------------------------------------------------------------------
\subsection{ROI definitions, noise-ceiling threshold, and aggregation}
\label{app:rois}

The exact ROI masks and voxel counts for each subject are reported in
Table~\ref{tab:voxels}. Places combines OPA, PPA, and RSC; Faces combines OFA
and FFA-1/2; Bodies combines EBA and FBA-1/2; and EBA is also analyzed
separately. EBA comprises approximately 80\% of the voxels in the Bodies mask,
so Bodies and EBA are not treated as independent ROIs in pairwise comparisons.

% ===== Table: voxels per ROI and subject (tab:voxels) =====
% Included from neurips_2025.tex -- do not compile on its own.

\begin{table}[!h]
  \centering
  \caption{Voxels per ROI and subject. All models were fit on 10{,}000 images per
  subject (3 repeats each, responses averaged per image). ``Places'' merges OPA, PPA
  and RSC; EBA is a subset of the body-selective mask, so the late group contains
  three independent ROIs. V1 and V4 were fit in a separate layer-matched leg
  (MPNet layers 2 and 6) and are never pooled with the late ROIs. Masks are NSD's \texttt{floc-places} labels 1--3, \texttt{floc-faces} labels 1--3, \texttt{floc-bodies} labels 1--3 (EBA is label 1 alone) and \texttt{prf-visualrois} (V1v+V1d and hV4), in each subject's 1.8\,mm functional space.}
  \label{tab:voxels}
  \begin{tabular}{lrr}
    \toprule
    ROI & subj01 & subj02 \\
    \midrule
    Places (OPA+PPA+RSC) & 2{,}493 & 2{,}593 \\
    Faces                & 908     & 917     \\
    Bodies               & 3{,}346 & 3{,}589 \\
    EBA ($\subset$ Bodies) & 2{,}694 & 2{,}776 \\
    \addlinespace
    V1                   & 1{,}291 & 920     \\
    V4                   & 568     & 414     \\
    \bottomrule
  \end{tabular}
\end{table}

Within each ROI, we retained voxels with noise-corrected SNR
(ncsnr) $\geq 0.2$. This threshold was chosen to stabilize noise-ceiling
normalization rather than to select voxels based on model prediction accuracy.
Prediction accuracy for each voxel is divided by its noise ceiling before
averaging across voxels (Section~\ref{subsec:model}); voxels with ceilings
close to zero can therefore produce unstable normalized correlations and
disproportionately influence the ROI mean.

The noise ceiling in correlation units is
\[
  c_v =
  \sqrt{\frac{\mathrm{ncsnr}_v^2}
  {\mathrm{ncsnr}_v^2 + 1/n}},
\]
with $n=3$ averaged repeats
\citep{allen2022massive,schoppe2016measuring}. Thus,
ncsnr $\geq 0.2$ corresponds to a noise ceiling of approximately $0.33$.
The same voxel threshold was applied to every model entering a contrast, and
no surviving voxel produced an undefined correlation in either the joint or
language-only model.

To test sensitivity to this threshold, we rescored the matched-length contrast
using voxels with ncsnr $\geq 0.3$ and $\geq 0.4$, with the same out-of-fold
predictions and bootstrap draws. In every category-selective ROI and subject,
the contrast changed by at most $0.003$, and every interval still excluded
zero. The original one-caption Places contrast changed by at most $0.002$ and
retained its sign and interval conclusion.

We also tested sensitivity to the summary statistic across voxels. In addition
to the unweighted mean used in the main analysis, we recomputed each ROI using
the regional median and a noise-ceiling-weighted mean. For voxel $v$, let
$r_v$ denote its unnormalized prediction correlation and $c_v$ its noise
ceiling. Weighting each normalized correlation $r_v/c_v$ by $c_v$ is
equivalent to
\begin{equation}
    \frac{\sum_v r_v}{\sum_v c_v}.
\end{equation}
Replacing the unweighted mean with either alternative preserved the sign and
interval conclusion of every matched-length contrast and changed no estimate
by more than $0.005$. The noise-ceiling-weighted mean was within $0.0015$ of
the unweighted mean for every ROI and subject. The scrambled-text control was
more sensitive in magnitude to aggregation, although its sign and its distinct
behavior in Places were preserved. Complete results are reported in
Table~\ref{tab:aggregation}.

% ===== Table: aggregator sensitivity of the floor (tab:aggregation) =====
% Included from neurips_2025.tex -- do not compile on its own.

\begin{table}[!htbp]
  \centering
  \small
  \caption{Aggregator sensitivity of the content-destroyed floor. Each ROI mean over
  voxels is recomputed three ways --- the unweighted mean of ceiling-normalized
  accuracy used throughout the paper, the regional median, and a mean weighting each
  voxel by its own noise ceiling. All values are full-sample point estimates. Every
  floor value keeps its sign, and every intact contrast keeps its sign, under all
  three aggregators. The
  argument of Section~\ref{subsec:floor} requires the floor to be at least as large
  as the intact contrast; $\dagger$ marks the two cells where it is not, both in
  subj02 and both under the median. Places is the region where the floor is opposite
  in sign to the intact contrast, so the magnitude comparison does not apply there;
  its floor stays positive under all three aggregators.}
  \label{tab:aggregation}
  \begin{tabular}{llrrrrrr}
    \toprule
    & & \multicolumn{3}{c}{Content-destroyed floor} & \multicolumn{3}{c}{Intact matched contrast} \\
    \cmidrule(lr){3-5}\cmidrule(lr){6-8}
    ROI & Subj & Mean & Median & Ceil.\ wt. & Mean & Median & Ceil.\ wt. \\
    \midrule
    Faces  & subj01 & $-0.062$ & $-0.052$ & $-0.067$ & $-0.040$ & $-0.040$ & $-0.041$ \\
    Faces  & subj02 & $-0.052$ & $-0.056$ & $-0.056$ & $-0.040$ & $-0.036$ & $-0.041$ \\
    \addlinespace
    Bodies & subj01 & $-0.081$ & $-0.086$ & $-0.085$ & $-0.042$ & $-0.041$ & $-0.042$ \\
    Bodies & subj02 & $-0.038$ & $-0.018^{\dagger}$ & $-0.041$ & $-0.034$ & $-0.032$ & $-0.035$ \\
    \addlinespace
    EBA    & subj01 & $-0.084$ & $-0.091$ & $-0.088$ & $-0.042$ & $-0.041$ & $-0.042$ \\
    EBA    & subj02 & $-0.045$ & $-0.027^{\dagger}$ & $-0.047$ & $-0.034$ & $-0.031$ & $-0.034$ \\
    \midrule
    Places & subj01 & $+0.017$ & $+0.028$ & $+0.021$ & $-0.031$ & $-0.034$ & $-0.031$ \\
    Places & subj02 & $+0.022$ & $+0.032$ & $+0.028$ & $-0.023$ & $-0.026$ & $-0.023$ \\
    \bottomrule
  \end{tabular}
\end{table}

% ---------------------------------------------------------------------
\subsection{Embedding implementation details}
\label{app:embeddings}

\paragraph{Vision embeddings.}
Across the two subjects, the NSD stimulus set contained 19{,}000 distinct
images: 9{,}000 unique images per subject and 1{,}000 shared images. Images
were processed using DINOv2 ViT-B/14
(\texttt{vit\_base\_patch14\_dinov2.lvd142m})
\citep{oquab2024dinov2}. Image preprocessing used
\texttt{create\_transform} from \texttt{timm} version 1.0.28.

Features were extracted from DINOv2 layers 2, 6, and 11:
\begin{enumerate}[topsep=0pt,parsep=0pt]
    \item layer 2 for V1: the $37\times37$ grid of patch tokens was reduced to
    a $4\times4$ grid by adaptive average pooling and flattened
    ($12{,}288$ features), preserving spatial position;
    \item layer 6 for V4: the CLS token was concatenated with the mean over
    patch tokens ($1{,}536$ features);
    \item layer 11 for the category-selective ROIs: pooled in the same way as
    layer 6 ($1{,}536$ features).
\end{enumerate}

These visual layers were paired with MPNet layers at approximately matched
encoder depth: layer 2 for V1, layer 6 for V4, and the final MPNet block
(layer 12) for the category-selective ROIs. Pairwise ROI comparisons were
performed only between ROIs using the same DINOv2--MPNet layer pair.

\paragraph{Text conditions.}
MS~COCO provides five independently written captions per image
\citep{lin2014microsoft}. Individual captions averaged approximately
10.4 words (range 6--47). Captions were used either individually or
concatenated within an image to construct four- and five-caption predictors.
Concatenation followed caption-identifier order, with captions separated by a
space and a terminal period added where one was missing. Four concatenated
captions averaged 41.7 words (range 31--91), and five captions averaged
approximately 52 words.

Localized Narratives are transcribed spoken descriptions from one annotator
per description \citep{pont2020connecting}. The selected narratives averaged
41.3 words (range 4--211). Their register was strongly formulaic: 85.5\% contained
``image'' or ``picture,'' 54.9\% contained ``see,'' nearly half began with a
variant of ``In this image I can see,'' and ``see'' was the most frequent
content word, accounting for 6.9\% of content tokens.

Four concatenated captions therefore approximately matched the average word
count of a full narrative, but not its annotator count, lexical composition,
redundancy, or discourse structure.

We also included two narrative truncations, each cut to the word count of that
image's selected caption (mean 10.4 words, range 4--47). Narrative-first kept
the opening words, whereas Narrative-span retained a contiguous interior span.
These conditions therefore matched caption length image by image rather than
only in the dataset mean.

Word-count matching also closely matched the number of subword tokens received
by MPNet. Individual captions averaged 1.134 MPNet tokens per word, narratives
1.144, and four concatenated captions 1.156. The matched four-caption and
narrative conditions therefore contained 48.1 and 47.0 MPNet tokens on
average, respectively, a difference of 2.3\%. No text approached MPNet's
384-token input limit, so no condition was truncated by the encoder.

All text conditions are summarized in Table~\ref{tab:tab_predictors}.

% ===== Table: the length ladder (tab:rungs) =====
% Included from neurips_2025.tex -- do not compile on its own.

\begin{table}[!h]
  \centering
  \caption{Text predictors used in the length-matching analysis. Word counts are dataset means. Four concatenated COCO captions approximately match the full Localized Narrative in length, but contain descriptions from four annotators. The two truncated-narrative conditions are each cut to the word count of that image's selected caption and differ only in where the segment is taken.}
  \label{tab:tab_predictors}
  \begin{tabular}{llrr}
    \toprule
    Predictor & Text input & $\sim$Words & Annotators \\
    \midrule
    COCO $\times$1 & one COCO caption                         & 10.4 & 1 \\
    COCO $\times$4 & four concatenated COCO captions          & 41.7 & 4 \\
    COCO $\times$5 & five concatenated COCO captions          & 52.0 & 5 \\
    \addlinespace
    Narrative      & full Localized Narrative                 & 41.3 & 1 \\
    Narrative-first & opening words, cut to caption length    & 10.4 & 1 \\
    Narrative-span  & interior span, cut to caption length    & 10.4 & 1 \\
    \bottomrule
  \end{tabular}
\end{table}

\paragraph{Text embeddings.}
Text embeddings were extracted from
\texttt{all-mpnet-base-v2}
\citep{song2020mpnet,reimers2019sentence}. Hidden states from MPNet layers 2,
6, and 12 were paired with the DINOv2 features used for V1, V4, and the
category-selective ROIs, respectively. These layers were fixed in advance to
match relative encoder depth; they were not selected based on neural
prediction performance.

For every MPNet layer, each text was represented by the mean of the
768-dimensional hidden states across all non-padding tokens, including the CLS
and separator tokens. Thus, all language predictors used the same
token-averaging procedure and differed only in their input text and, across
ROI groups, encoder depth.

All visual and language feature spaces were projected to 256 dimensions using
scikit-learn's \texttt{SparseRandomProjection}
\citep{achlioptas2003database}. Equalizing feature dimensionality prevents
differences between predictors from arising simply because one representation
contains more features \citep{conwell2024large}.

% ---------------------------------------------------------------------
\subsection{Scrambling and reconstruction controls}
\label{app:scrambling}

Scrambled text predictors were constructed using spaCy 3.8.14
\citep{honnibal2020spacy}. The pipeline was
\[
\text{raw text}
\rightarrow
\text{spaCy tokenization}
\rightarrow
\text{token substitution}
\rightarrow
\text{heuristic reconstruction}.
\]

We treated nouns, proper nouns, verbs, adjectives, and numerals as content
words. Each content word was replaced with a different word carrying the same
fine-grained part-of-speech tag whenever possible. When no alternative was
available, the procedure fell back first to the corresponding coarse
part-of-speech class and then to the complete content-word vocabulary.
Replacement words were sampled in proportion to their frequency in the same
dataset and could not be content words appearing in the original description
for that image. Copular forms of \emph{be}, including ``be,'' ``are,'' and
``is,'' were preserved.

For example, a caption reading ``A black dog runs under a blue sky'' could
become ``A silver table jumps under a green sea.'' The manipulation preserved
text length, function words, part-of-speech composition, sentence
segmentation, content-word counts, and content-word positions while removing
the original image-specific content words.

Because replacements were sampled from the dataset frequency distribution
independently of the image, dataset-typical vocabulary was redistributed
across images rather than removed. For example, a person word
(\emph{man}, \emph{woman}, \emph{people}, \emph{person}, \emph{girl},
\emph{boy}, \emph{child}, and plurals) occurred in 33\% of intact single
captions and 28\% of scrambled ones, in 44\% versus 70\% of four-caption
texts, and in 47\% versus 51\% of narratives. The identity of a substituted
content word therefore carried no image-specific information.

Because spaCy tokenization followed by heuristic reconstruction could itself
modify the text, we constructed a reconstruction control. Each description was
tokenized and reconstructed using the same pipeline but without substitution.
Replacing the original texts with reconstructed versions at the one-caption
rung changed $\Delta_V(L)$ by at most $1.2\times10^{-4}$ in any
category-selective ROI or subject. The reconstructed caption--narrative
contrast matched the intact contrast to four decimal places
(Places: $+0.0118/{+0.0146}$ reconstructed versus
$+0.0118/{+0.0147}$ intact). Tokenization and reconstruction therefore
contributed no measurable effect.

\begin{figure}[t]
  \centering
  \includegraphics[
    width=\linewidth,
    alt={Two horizontal grouped bar charts comparing COCO captions in blue and
    Localized Narratives in orange across seven word classes. Panel a shows
    tokens per description: narratives contain more tokens from nearly every
    class because they are longer. Panel b shows the same classes as proportions
    of tokens: captions contain higher proportions of nouns, adjectives, color
    words, numbers, and multi-word spatial phrases, whereas narratives contain
    higher proportions of verbs and single-word spatial prepositions.}
  ]{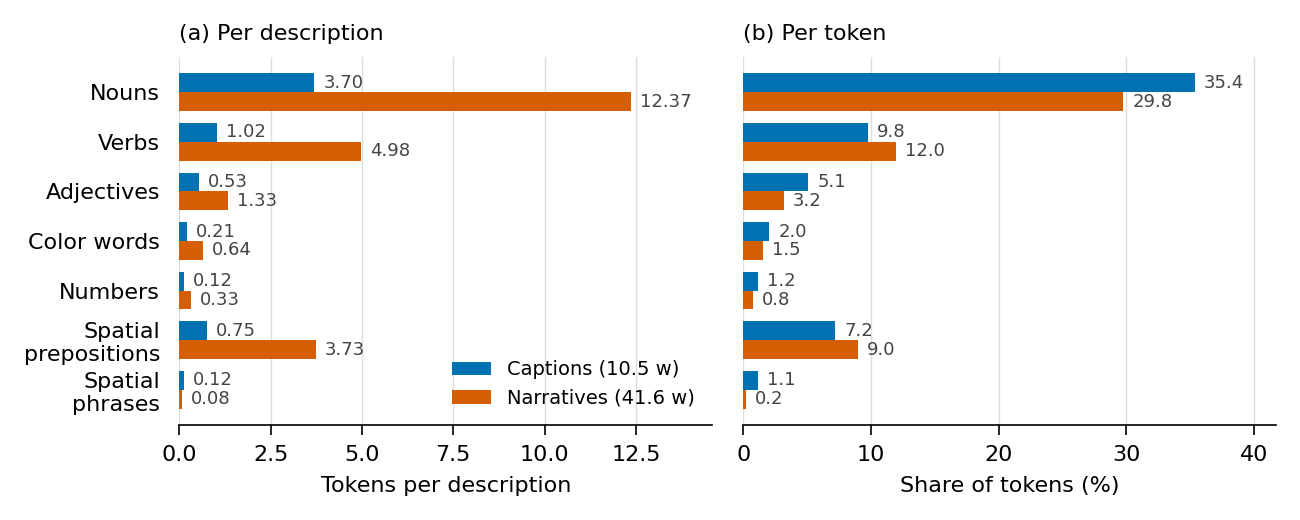}
  \caption{spaCy transformer parse of both datasets (73{,}000 images each).
  \textbf{(a)} Tokens per description by word class. Narratives average
  41.6 words compared with 10.5 for captions and contain more tokens from
  nearly every class.
  \textbf{(b)} The same counts as proportions of tokens. After accounting for
  length, captions contain higher proportions of nouns, adjectives, color
  words, numbers, and multi-word spatial phrases, whereas narratives contain
  higher proportions of verbs and single-word spatial prepositions.}
  \label{fig:corpus}
\end{figure}

The two text datasets also differ in linguistic composition.
Figure~\ref{fig:corpus} summarizes the spaCy parse both as counts per
description and as proportions of tokens.

Substitution and scrambling controls have a long history in studies
dissociating lexical-semantic information from other linguistic properties
\citep{miller1963some,fedorenko2010new,mollica2020composition,
lerner2011topographic}. Related controls are used in NLP to determine what
information probes and learned representations capture
\citep{hewitt2019designing,sinha2021masked,oconnor2021context,
yuksekgonul2023when}. In a closely related neural setting,
\citet{kauf2024lexical} used sentence perturbations to identify stimulus
properties contributing to model-to-brain similarity in the language network.

% ---------------------------------------------------------------------
\subsection{Banded ridge regression}
\label{app:ridge}

Given feature bands $X_1,\dots,X_B$, each z-scored using statistics from the
training fold, band $b$ received its own regularization penalty $\alpha_b$.
Weights were estimated as
\begin{equation}
\hat{W}
=
\arg\min_W
\left\| Y-\sum_b X_bW_b\right\|_2^2
+
\sum_b \alpha_b\| W_b\|_2^2 .
\end{equation}

We implemented the solution in closed form by rescaling each band as
\begin{equation}
    \tilde{X}_b = X_b/\sqrt{\alpha_b},
\end{equation}
concatenating the scaled bands, and solving the standard ridge normal
equations with unit penalty. This yields predictions identical to the
corresponding band-specific ridge formulation.

Each $\alpha_b$ was selected from 18 logarithmically spaced values spanning
$10^{-1}$ to $10^{16}$. For models containing multiple searched bands, the
candidate set was the Cartesian product of these values. The implementation
caps the search at 400 candidate combinations and coarsens the grid while
preserving its endpoints if this cap is exceeded. The present analyses contain
at most two searched bands, so all $18^2=324$ combinations were evaluated and
the cap never applied.

Regularization was selected independently for each voxel using nested
cross-validation. The data were divided into five outer folds. Within each
outer fold, the training data were split 80/20 into inner training and
validation sets. Every candidate penalty combination was fit on the inner
training set, and the combination maximizing Pearson correlation on the inner
validation set was selected for that voxel. The selected combination was then
refit on the complete outer-training fold and used to predict the held-out
outer fold.

% ---------------------------------------------------------------------
\subsection{Prediction metric and contrast decomposition}
\label{app:metric}

Prediction accuracy $r$ was computed separately for each voxel as the Pearson
correlation between concatenated out-of-fold predictions and repeat-averaged
fMRI responses. This correlation was divided by the voxel's noise ceiling
$c_v$ defined above, and the resulting normalized correlations were averaged
across voxels within each ROI.

For each language predictor $L$, we fit a joint vision--language model and a
language-only model. The additional predictive contribution of vision is
\begin{equation}
    \Delta_V(L)=r(V{+}L)-r(L),
\end{equation}
which is a difference in correlation, not variance explained.

For two language predictors $L_A$ and $L_B$,
\begin{equation}
\Delta_V(L_A)-\Delta_V(L_B)
=
\left[r(V{+}L_A)-r(V{+}L_B)\right]
-
\left[r(L_A)-r(L_B)\right].
\end{equation}
The first bracket is the change in joint vision--language prediction and the
second is the change in language-only prediction. Their difference is exactly
the observed contrast.

Figure~\ref{fig:main}C shows this decomposition for the original and
matched-length Places comparisons. Table~\ref{tab:decomposition} reports both
terms for every ROI and subject. Absolute accuracies, $r(V{+}L)$ and $r(L)$,
for every text condition, ROI, and subject are reported in
Table~\ref{tab:results}.

% ===== Table: decomposition of the contrast into joint and language-only terms (tab:decomposition) =====
% Generated by decomp_table.py from null_gate, cocoC4s_gate and the layer-matched V1/V4 runs.

\begin{table}[!htbp]
  \centering
  \small
  \caption{Decomposition of the caption--narrative contrast (Eq.~\ref{eq:contrast}) for every
  ROI and subject. Joint is $r(V{+}L)_{\mathrm{cap}}-r(V{+}L)_{\mathrm{nar}}$, Lang is
  $r(L)_{\mathrm{cap}}-r(L)_{\mathrm{nar}}$, and the contrast is Joint minus Lang. Original
  compares one caption with the narrative; matched compares four concatenated captions with
  the narrative. Full-sample values. V1 and V4 use their own encoder layer pairs.}
  \label{tab:decomposition}
  \begin{tabular}{llrrrrrr}
    \toprule
    & & \multicolumn{3}{c}{Original (1 caption)} & \multicolumn{3}{c}{Matched (4 captions)} \\
    \cmidrule(lr){3-5} \cmidrule(lr){6-8}
    ROI & Subj & Joint & Lang & Contrast & Joint & Lang & Contrast \\
    \midrule
    Places & subj01 & $+0.0008$ & $-0.0110$ & $+0.0118$ & $+0.0181$ & $+0.0488$ & $-0.0307$ \\
    Places & subj02 & $-0.0003$ & $-0.0149$ & $+0.0147$ & $+0.0146$ & $+0.0378$ & $-0.0232$ \\
    \addlinespace
    Faces & subj01 & $+0.0048$ & $+0.0124$ & $-0.0076$ & $+0.0240$ & $+0.0643$ & $-0.0403$ \\
    Faces & subj02 & $+0.0054$ & $+0.0129$ & $-0.0075$ & $+0.0190$ & $+0.0591$ & $-0.0401$ \\
    \addlinespace
    Bodies & subj01 & $+0.0097$ & $+0.0207$ & $-0.0110$ & $+0.0336$ & $+0.0753$ & $-0.0417$ \\
    Bodies & subj02 & $+0.0069$ & $+0.0108$ & $-0.0039$ & $+0.0251$ & $+0.0595$ & $-0.0344$ \\
    \addlinespace
    EBA & subj01 & $+0.0107$ & $+0.0225$ & $-0.0118$ & $+0.0344$ & $+0.0763$ & $-0.0419$ \\
    EBA & subj02 & $+0.0073$ & $+0.0101$ & $-0.0028$ & $+0.0257$ & $+0.0597$ & $-0.0340$ \\
    \addlinespace
    V1 & subj01 & $+0.0004$ & $-0.0083$ & $+0.0087$ & $+0.0017$ & $+0.0400$ & $-0.0383$ \\
    V1 & subj02 & $+0.0005$ & $-0.0031$ & $+0.0035$ & $+0.0025$ & $+0.0488$ & $-0.0463$ \\
    \addlinespace
    V4 & subj01 & $+0.0019$ & $+0.0209$ & $-0.0190$ & $+0.0046$ & $+0.0662$ & $-0.0616$ \\
    V4 & subj02 & $+0.0016$ & $+0.0178$ & $-0.0162$ & $+0.0040$ & $+0.0596$ & $-0.0556$ \\
    \bottomrule
  \end{tabular}
\end{table}

% ===== Table: full-sample r(V+L), r(L), uniqueV per text condition, ROI, subject (tab:results) =====
% Generated by results_table.py.

\begin{table}[!htbp]
  \centering
  \small
  \caption{Prediction accuracy of every text predictor. For each ROI, subject, and text
  condition: the joint vision--language model $r(V{+}L)$, the language-only model $r(L)$, and
  their difference uniqueV (Eq.~\ref{eq:uniquev}), all ceiling-normalized, averaged over voxels,
  and computed on the full sample. Each ROI is fit at its own encoder layer pair. The scrambled
  single caption was not fit at the V1 and V4 layer pairs. The five-caption and 10-word
  narrative conditions were run in Places only; the random-subsample conditions are in
  Table~\ref{tab:subsample}.}
  \label{tab:results}
  \begin{tabular}{llrrrrrr}
    \toprule
    & & \multicolumn{3}{c}{subj01} & \multicolumn{3}{c}{subj02} \\
    \cmidrule(lr){3-5} \cmidrule(lr){6-8}
    ROI & Text predictor & $r(V{+}L)$ & $r(L)$ & uniqueV & $r(V{+}L)$ & $r(L)$ & uniqueV \\
    \midrule
    Places & 1 caption & $0.642$ & $0.546$ & $0.095$ & $0.666$ & $0.573$ & $0.093$ \\
     & 1 caption, scrambled & $0.621$ & $0.091$ & $0.531$ & $0.648$ & $0.103$ & $0.545$ \\
     & 4 captions & $0.659$ & $0.606$ & $0.053$ & $0.681$ & $0.626$ & $0.055$ \\
     & 5 captions & $0.662$ & $0.612$ & $0.049$ & $0.683$ & $0.631$ & $0.052$ \\
     & 4 captions, scrambled & $0.621$ & $0.195$ & $0.426$ & $0.648$ & $0.218$ & $0.430$ \\
     & Narrative & $0.641$ & $0.557$ & $0.084$ & $0.666$ & $0.588$ & $0.078$ \\
     & Narrative-first (10 words) & $0.625$ & $0.409$ & $0.216$ & $0.652$ & $0.443$ & $0.208$ \\
     & Narrative-span (10 words) & $0.624$ & $0.387$ & $0.237$ & $0.651$ & $0.423$ & $0.227$ \\
     & Narrative, scrambled & $0.622$ & $0.213$ & $0.409$ & $0.648$ & $0.240$ & $0.408$ \\
    \addlinespace
    Faces & 1 caption & $0.580$ & $0.506$ & $0.074$ & $0.540$ & $0.459$ & $0.081$ \\
     & 1 caption, scrambled & $0.557$ & $0.116$ & $0.441$ & $0.519$ & $0.102$ & $0.417$ \\
     & 4 captions & $0.600$ & $0.558$ & $0.042$ & $0.553$ & $0.505$ & $0.048$ \\
     & 4 captions, scrambled & $0.558$ & $0.217$ & $0.341$ & $0.519$ & $0.183$ & $0.336$ \\
     & Narrative & $0.576$ & $0.494$ & $0.082$ & $0.534$ & $0.446$ & $0.088$ \\
     & Narrative, scrambled & $0.557$ & $0.154$ & $0.402$ & $0.519$ & $0.131$ & $0.388$ \\
    \addlinespace
    Bodies & 1 caption & $0.680$ & $0.617$ & $0.063$ & $0.669$ & $0.604$ & $0.065$ \\
     & 1 caption, scrambled & $0.644$ & $0.167$ & $0.477$ & $0.636$ & $0.163$ & $0.473$ \\
     & 4 captions & $0.704$ & $0.672$ & $0.032$ & $0.687$ & $0.653$ & $0.035$ \\
     & 4 captions, scrambled & $0.647$ & $0.309$ & $0.338$ & $0.638$ & $0.282$ & $0.355$ \\
     & Narrative & $0.670$ & $0.596$ & $0.074$ & $0.662$ & $0.593$ & $0.069$ \\
     & Narrative, scrambled & $0.645$ & $0.225$ & $0.419$ & $0.636$ & $0.242$ & $0.394$ \\
    \addlinespace
    EBA & 1 caption & $0.686$ & $0.624$ & $0.062$ & $0.664$ & $0.599$ & $0.065$ \\
     & 1 caption, scrambled & $0.648$ & $0.174$ & $0.475$ & $0.630$ & $0.169$ & $0.461$ \\
     & 4 captions & $0.710$ & $0.678$ & $0.032$ & $0.682$ & $0.648$ & $0.034$ \\
     & 4 captions, scrambled & $0.651$ & $0.318$ & $0.334$ & $0.632$ & $0.290$ & $0.342$ \\
     & Narrative & $0.675$ & $0.602$ & $0.074$ & $0.657$ & $0.589$ & $0.068$ \\
     & Narrative, scrambled & $0.649$ & $0.231$ & $0.418$ & $0.631$ & $0.243$ & $0.387$ \\
    \addlinespace
    V1 & 1 caption & $0.624$ & $0.198$ & $0.426$ & $0.629$ & $0.245$ & $0.384$ \\
     & 4 captions & $0.626$ & $0.247$ & $0.379$ & $0.631$ & $0.297$ & $0.334$ \\
     & 4 captions, scrambled & $0.624$ & $0.091$ & $0.533$ & $0.629$ & $0.141$ & $0.488$ \\
     & Narrative & $0.624$ & $0.207$ & $0.417$ & $0.629$ & $0.248$ & $0.381$ \\
     & Narrative, scrambled & $0.624$ & $0.077$ & $0.547$ & $0.628$ & $0.099$ & $0.529$ \\
    \addlinespace
    V4 & 1 caption & $0.541$ & $0.323$ & $0.218$ & $0.566$ & $0.339$ & $0.227$ \\
     & 4 captions & $0.544$ & $0.369$ & $0.175$ & $0.568$ & $0.381$ & $0.188$ \\
     & 4 captions, scrambled & $0.540$ & $0.185$ & $0.355$ & $0.565$ & $0.220$ & $0.345$ \\
     & Narrative & $0.540$ & $0.303$ & $0.237$ & $0.564$ & $0.321$ & $0.243$ \\
     & Narrative, scrambled & $0.539$ & $0.127$ & $0.413$ & $0.564$ & $0.133$ & $0.431$ \\
    \bottomrule
  \end{tabular}
\end{table}

% ---------------------------------------------------------------------
\subsection{Bootstrap inference}
\label{app:bootstrap}

Each subject was evaluated using out-of-fold predictions for all 10{,}000
images viewed by that subject. The subjects share only 1{,}000 images, so their
agreement largely reflects replication across distinct stimulus sets.

Uncertainty was estimated independently for each subject using a paired
bootstrap over evaluation images
\citep{efron1979bootstrap,efron1994introduction}. For each of 1{,}000 draws,
10{,}000 images were sampled with replacement. The same sampled image indices
were applied simultaneously to the observed fMRI responses and to the
predictions from every model entering a contrast, after which the contrast was
recomputed. Models were not refit during bootstrapping; all predictions had
already been obtained out of fold.

We report 95\% percentile intervals defined by the 2.5th and 97.5th
percentiles of the bootstrap distribution. Point estimates are full-sample
values, and the bootstrap is used only to estimate intervals. The mean of the
bootstrap distribution differed from the full-sample estimate by less than
$10^{-4}$ for every reported contrast.

Because only two subjects were analyzed, we make no population-level inference
\citep{naselaris2021extensive}. Effects are reported separately for each
subject.

% ---------------------------------------------------------------------
\subsection{Text-selection robustness}
\label{app:text_redraw}

The one-caption versus one-narrative comparison could depend on which available
text was selected for an image. We therefore repeated the comparison using
five independent redraws while holding the fMRI responses, visual predictor,
regularization grid, random seed, cross-validation folds, and bootstrap
indices fixed \citep{steegen2016increasing}. Only the selected texts changed.

For each image, one caption and one narrative were selected from the available
texts by a keyed hash of the image identifier. The original predictors used
one key, and each redraw used a different key. MS~COCO provides approximately
five captions per image, so a redraw changed the caption for about 80\% of
images. Localized Narratives provides only about 1.16 narratives per image on
average, so only about 3.7\% of narratives changed. The redraw therefore tests
caption selection much more strongly than narrative selection.

Across the original selection and five additional redraws, the
caption--narrative contrast in Places remained positive and its bootstrap
interval excluded zero in both subjects. Across the six selections, the
contrast ranged from $+0.0068$ to $+0.0136$ in subj01 and from
$+0.0122$ to $+0.0158$ in subj02. We therefore interpret the direction of the
original contrast rather than its exact magnitude as the robust result.

% ---------------------------------------------------------------------
\subsection{Shared-stimulus sensitivity analysis}
\label{app:shared_stimuli}

The two subjects viewed 1{,}000 images in common. To test whether agreement
between subjects depended on their largely distinct stimulus sets, we rescored
the matched-length comparison using only these shared images while leaving the
training data unchanged.

The negative matched-length contrast was retained in Faces
($-0.030/{-0.032}$), Bodies ($-0.031/{-0.023}$), and EBA
($-0.032/{-0.022}$), with bootstrap intervals excluding zero in both subjects.
In Places, the contrasts were
$-0.0127$ $[-0.0245,-0.0003]$ in subj01 and
$-0.0073$ $[-0.0199,+0.0045]$ in subj02. Intervals were substantially wider
with one tenth of the evaluation images, and the shared-stimulus analysis
therefore provides weaker support for the Places effect than for the other
category-selective ROIs.

% ---------------------------------------------------------------------
\subsection{Additional scramble analyses}
\label{app:scramble_extra}

At the original, unmatched text lengths, the scrambled caption--narrative
comparison in Places was $+0.1224$ in subj01 and $+0.1375$ in subj02.
Approximately matching text length reduced this contrast by about 85\%.

One possible source of difference between the intact predictors is that the
four-caption condition combines descriptions from four annotators, which may
repeatedly name image content that appears only once in the single-annotator
narrative. Scrambling removes the original content words while retaining this
and other differences between the text conditions, so it cannot isolate
annotator redundancy from other dataset properties.

To test whether description-level counts alone contain predictive signal, we
fit individual count features from the narratives as the sole language
predictor in Places, with the visual band and all other settings unchanged.
The number of determiners reached $r(L)=0.142/0.161$, the number of pronouns
$0.130/0.152$, and the number of spatial prepositions $0.115/0.141$, compared
with $0.557/0.588$ for the intact narrative embedding. Thus, even a single
description-level count with no information about which particular words
occurred can recover a substantial fraction of the language predictor's
accuracy.

% ---------------------------------------------------------------------
\subsection{Random-subsample control for the length effect}
\label{app:subsample}

To test whether the length effect in Figure~\ref{fig:main}A depended on which
words were removed, we constructed random $k$-word subsamples of the narrative
for each image, with $k=10$, $20$, and $41$. Words were drawn without
replacement, their original order was preserved, and one draw per image was
generated using a keyed hash of the image identifier. Because 58\% of
narratives contain at most 41 words, the $k=41$ condition leaves those
narratives intact and approximates the full text.

Subsampled texts were embedded using the same MPNet pipeline as every other
language predictor and fit in Places with all settings fixed to the
configuration of the original comparison.

\begin{table}[!htbp]
  \centering
  \small
  \caption{Random-subsample control in Places. $\Delta_V(L)$ for random
  $k$-word narrative subsamples and the full narrative, with paired-bootstrap
  contrasts between adjacent conditions. Every reported contrast excludes
  zero in both subjects.}
  \label{tab:subsample}
  \begin{tabular}{lrr}
    \toprule
    Predictor & subj01 & subj02 \\
    \midrule
    Narrative, random 10 words & $0.248$ & $0.239$ \\
    Narrative, random 20 words & $0.145$ & $0.144$ \\
    Narrative, random 41 words & $0.093$ & $0.092$ \\
    Narrative, full            & $0.084$ & $0.078$ \\
    \addlinespace
    10 words minus 20 words
      & $+0.103$ $[+0.095,+0.110]$
      & $+0.095$ $[+0.087,+0.103]$ \\
    20 words minus 41 words
      & $+0.053$ $[+0.048,+0.057]$
      & $+0.052$ $[+0.047,+0.057]$ \\
    41 words minus full
      & $+0.009$ $[+0.007,+0.011]$
      & $+0.014$ $[+0.012,+0.016]$ \\
    \bottomrule
  \end{tabular}
\end{table}

$\Delta_V(L)$ fell monotonically with $k$ in both subjects
(Table~\ref{tab:subsample}). The random 10-word subsample
($0.248/0.239$) lies in the same regime as the two contiguous caption-length
narrative cuts in the main analysis ($0.216$--$0.237$). The two analyses were
fit separately, so this cross-run comparison is indicative only, but it shows
that the short-text result did not depend on selecting the opening or an
interior contiguous span.

The small residual gap between the 41-word subsample and full narrative is
carried by the 42\% of narratives longer than 41 words that were actually
subsampled; the remaining texts are identical. In the 10-word condition, the
joint model selected the upper edge of the regularization grid for 9.2\% of
voxels in subj01 and 5.9\% in subj02, and for at most 2.3\% in the other
conditions, well below the 30\% level at which we would treat the grid as
binding.

The pattern is consistent with a variance-reduction or content-coverage
account of the mean-pooled text embedding: as more words are sampled, the
language predictor becomes stronger and leaves less additional predictive
contribution for vision. With a single draw per image, however, reduced
sampling noise and increased content coverage cannot be separated.

\paragraph{Pooled 10-word draws.}
To distinguish long encoder inputs from repeated short views, we constructed
seven additional independent random 10-word draws per image (seeds 1--7; the
preceding 10-word condition used seed 0). Each draw was embedded separately,
and the resulting embeddings were averaged over $m=1$, $2$, $4$, and $8$
draws. Pools were nested, so moving from $m$ to $2m$ only added new draws.
No encoder forward pass saw more than 10 words.

The pooled predictors were fit in Places alongside the full narrative with all
other settings fixed as above.

\begin{table}[!htbp]
  \centering
  \small
  \caption{Pooled-draws test in Places. $\Delta_V(L)$ for the mean of $m$
  independent random 10-word narrative embeddings, with the full narrative
  refit in the same run. Every pairwise contrast among the five predictors
  excludes zero in both subjects.}
  \label{tab:pooled}
  \begin{tabular}{lrr}
    \toprule
    Predictor & subj01 & subj02 \\
    \midrule
    Mean of 1 draw  & $0.248$ & $0.239$ \\
    Mean of 2 draws & $0.187$ & $0.180$ \\
    Mean of 4 draws & $0.143$ & $0.141$ \\
    Mean of 8 draws & $0.120$ & $0.116$ \\
    Narrative, full & $0.084$ & $0.078$ \\
    \addlinespace
    1 draw minus 8 draws
      & $+0.128$ $[+0.121,+0.134]$
      & $+0.122$ $[+0.116,+0.129]$ \\
    4 draws minus 8 draws
      & $+0.024$ $[+0.021,+0.026]$
      & $+0.025$ $[+0.022,+0.028]$ \\
    Full minus 8 draws
      & $-0.036$ $[-0.040,-0.032]$
      & $-0.038$ $[-0.042,-0.034]$ \\
    \bottomrule
  \end{tabular}
\end{table}

$\Delta_V(L)$ fell monotonically with $m$
(Table~\ref{tab:pooled}). Averaging eight 10-word views closed 78.0\% of the
gap between one 10-word view and the full narrative in subj01 and 76.2\% in
subj02. The step from four to eight draws still excluded zero, so these values
are lower bounds on the fraction recoverable with pooled short views.

A residual remained between the eight-view predictor and the full narrative
($-0.036/{-0.038}$ for full minus eight views), excluding zero in both
subjects. Thus, most of the length effect could be recovered without any
single encoder pass seeing a long input, but pooled short views did not fully
reproduce the full-narrative representation.

The interpretation is limited by increasing content coverage: eight
independent 10-word draws jointly cover an expected
$1-(31/41)^8\approx89\%$ of the words in a 41-word narrative. Pooling therefore
reduces sampling variability while also exposing the aggregate representation
to more of the narrative's content.

Upper-edge selection of the regularization grid in the joint model peaked for
a single draw (9.2\%/5.9\%) and declined with $m$
(3.9\%/2.6\% at $m=2$, 1.9\%/1.7\% at $m=4$, and
1.4\%/1.2\% at $m=8$). As an indicative cross-run comparison, averaging four
10-word draws ($0.143/0.141$) yielded almost the same $\Delta_V(L)$ as a
single random 20-word subsample ($0.145/0.144$).

% ---------------------------------------------------------------------
\subsection{Per-image dose-response of the length effect}
\label{app:dose}

The comparisons above contrast discrete text conditions that can differ along
more than one dimension. Narrative length also varies naturally across images
(range 4--211 words), allowing us to ask within a single predictor whether
images with longer descriptions leave less additional predictive contribution
for vision.

With each voxel's response and prediction z-scored over the evaluation set,
\[
  \mathrm{mean}_i(z_Y-z_P)^2 = 2-2r
\]
exactly. We therefore define the per-image quantity
\begin{equation}
  \delta_i =
  \mathrm{mean}_v\!\left[
  \frac{1}{2}
  \frac{
    (z_{Y,iv}-z_{L,iv})^2
    -
    (z_{Y,iv}-z_{V\!+\!L,iv})^2
  }{c_v}
  \right].
\end{equation}
Averaging $\delta_i$ over images exactly recovers $\Delta_V(L)$ for that
predictor; this equality was verified against each run's aggregate output.

For the full-narrative predictor, we correlated $\delta_i$ with the word count
of the corresponding narrative using cached out-of-fold predictions and the
same bootstrap draws as elsewhere. We also fit a standardized regression
including word count, type-token ratio, content-word repetition rate, and the
L2 norm of the sentence embedding.

\begin{table}[!htbp]
  \centering
  \small
  \caption{Per-image dose-response within the full-narrative predictor.
  Pearson correlation between $\delta_i$ and narrative word count over each
  subject's 10{,}000 images, and mean $\delta_i$ in the shortest and longest
  word-count deciles. Values are subj01/subj02. V1 and V4 use their own
  encoder layer pairs.}
  \label{tab:dose}
  \begin{tabular}{lrrr}
    \toprule
    ROI & $r(\delta_i,\text{words})$
        & shortest decile & longest decile \\
    \midrule
    Places & $-0.056/{-0.046}$ & $+0.17/{+0.15}$ & $+0.05/{+0.05}$ \\
    Faces  & $-0.077/{-0.093}$ & $+0.14/{+0.17}$ & $+0.04/{+0.02}$ \\
    Bodies & $-0.082/{-0.078}$ & $+0.13/{+0.14}$ & $+0.04/{+0.04}$ \\
    EBA    & $-0.078/{-0.078}$ & $+0.13/{+0.13}$ & $+0.05/{+0.04}$ \\
    \addlinespace
    V4     & $-0.121/{-0.104}$ & $+0.61/{+0.54}$ & $+0.18/{+0.17}$ \\
    V1     & $-0.201/{-0.211}$ & $+0.87/{+0.86}$ & $+0.30/{+0.24}$ \\
    \bottomrule
  \end{tabular}
\end{table}

Longer narratives left less additional predictive contribution for vision in
every ROI and both subjects, and $\delta_i$ declined monotonically across
word-count deciles (Table~\ref{tab:dose}). The effect was strongest in early
visual cortex, where the language predictor was weakest.

The same relation appeared in the differenced matched-length comparison.
Across images, the four-caption minus narrative difference in $\delta_i$
correlated with the difference in word count at $-0.10$ to $-0.21$ in every
ROI, with every bootstrap interval excluding zero. Images for which the caption
text was relatively longer therefore tended to show a relatively smaller
caption $\Delta_V(L)$.

Three qualifications limit the interpretation. First, word count and
type-token ratio are collinear within narratives because longer texts tend to
exhaust new word types. When both entered the regression, the slope shifted
toward type-token ratio and the standardized word-count coefficient became
null or slightly positive. The dose-response therefore does not isolate word
count from lexical variety.

Second, embedding norm carried no slope in any ROI or subject; every bootstrap
interval contained zero. The length effect is therefore not explained by
embedding magnitude.

Third, the relation was corpus-specific. Length of a single caption predicted
nothing, and in category-selective ROIs longer four-caption texts left slightly
more additional contribution for vision rather than less
(correlations $+0.03$ to $+0.05$), whereas V1 and V4 showed the same direction
as the narratives. Description length therefore covaries with
$\Delta_V(L)$ within the narrative predictor but is not a corpus-independent
law.

\end{document}